%% file: paper.tex
\documentclass[letterpaper]{article}
\input{common}

\title{%
    Poison Once, Control Anywhere:
    Clean-Text Visual Backdoors
    in \mbox{VLM-based Mobile Agents}}
\author{
  Xuan Wang\textsuperscript{\rm 1},
  Siyuan Liang\textsuperscript{\rm 2},
  Zhe Liu\textsuperscript{\rm 3},
  Yi Yu\textsuperscript{\rm 2},
  Aishan Liu\textsuperscript{\rm 4},
  Yuliang Lu\textsuperscript{\rm 1},
  Xitong Gao\textsuperscript{\rm 3},
  Ee-Chien Chang\textsuperscript{\rm 5}
}
\affiliations{
    \textsuperscript{\rm 1}National University of Defense Technology, China
    \textsuperscript{\rm 2}Nanyang Technological University, Singapore\\
    \textsuperscript{\rm 3}Chinese Academy of Sciences, China
    \textsuperscript{\rm 4}Beihang University, Beijing, China\\
    \textsuperscript{\rm 5}National University of Singapore, Singapore\\
    \{wangxuan21d\}@nudt.edu.cn,
    pandaliang521@gmail.com,
    liuzhe181@mails.ucas.edu.cn,
    yuyi0010@e.ntu.edu.sg,
    liuaishan@buaa.edu.cn,
    publicLuYL@126.com,
    xt.gao@siat.ac.cn,
    changec@comp.nus.edu.sg
}

\nocopyright
\begin{document}

\maketitle

\input{abstract}
\input{intro}
\input{related}
\input{figures/overview}
\input{threat}
\input{method}
\input{results}
\input{conclusion}

\bibliography{aaai2026}

\clearpage
\appendix
\setcounter{secnumdepth}{2}
\input{appendix}

\end{document}

%% file: common.tex
\usepackage{aaai2026}  
\usepackage{times}  
\usepackage{helvet}  
\usepackage{courier}  
\usepackage[hyphens]{url}  
\usepackage{graphicx} 
\usepackage{natbib}  
\usepackage{caption} 
\usepackage{algorithm}
\usepackage{algorithmic}
\usepackage{amsmath}
\usepackage{bm}
\usepackage{amsfonts}
\usepackage{amsmath,amssymb,amsfonts}
\usepackage{mathtools}
\usepackage{graphicx}
\usepackage{textcomp}
\usepackage{xcolor}
\usepackage{booktabs}
\usepackage{pifont}
\usepackage{multirow}
\usepackage{url}
\usepackage{cleveref}
\usepackage{subcaption}

\makeatletter

\DeclareRobustCommand\onedot{\futurelet\@let@token\@onedot}
\def\@onedot{\ifx\@let@token.\else.\null\fi}
\newcommand{\eg}{\emph{e.g\@\onedot}}
\newcommand{\etc}{\emph{etc\@\onedot}}
\newcommand{\etal}{\emph{et~al\@\onedot}}
\newcommand{\ie}{\emph{i.e\@\onedot}}

\DeclarePairedDelimiter{\norm}{\lVert}{\rVert}
\DeclarePairedDelimiter{\parens}{\lparen}{\rparen}

\DeclarePairedDelimiter{\bracks}{[}{]}
\DeclarePairedDelimiter{\braces}{\{}{\}}

\newcommand{\loss}{\mathcal{L}}

\newcommand{\tlbox}[1]{\begin{tabular}[c]{@{}l@{}}#1\end{tabular}}

\newcommand{\Method}{VIBMA}
\newcommand{\weight}{{\bm\theta}}
\newcommand{\model}{f_\weight}
\newcommand{\realset}{\mathbb{R}}
\newcommand{\dataset}{\mathcal{D}}
\newcommand{\dtrain}{\dataset_{\text{train}}}
\newcommand{\dclean}{\dataset_{\text{clean}}}
\newcommand{\dpoison}{\dataset_{\text{poison}}}
\newcommand{\noise}{\bm\delta}

\newcommand{\rate}{\gamma}
\newcommand{\image}{\textbf{x}}
\newcommand{\imageset}{\mathcal{X}}
\newcommand{\imagetriggered}{\image^{\textrm{triggered}}}
\newcommand{\imagetarget}{\image^{\textrm{target}}}
\newcommand{\imagepoison}{\image^{\textrm{poison}}}
\newcommand{\prompt}{\textbf{t}}
\newcommand{\promptset}{\mathcal{T}}
\newcommand{\thelabel}{y}
\newcommand{\labeltarget}{\thelabel^{\textrm{target}}}
\newcommand{\actionset}{\mathcal{A}}
\newcommand{\action}{\textbf{a}}
\newcommand{\actiontarget}{\action^{\textrm{target}}}
\newcommand{\reasonset}{\mathcal{C}}
\newcommand{\reason}{\textbf{c}}
\newcommand{\reasontarget}{\reason^{\textrm{target}}}
\newcommand{\mask}{\textbf{m}}
\newcommand{\trigger}{{\bm\tau}}
\newcommand{\runaction}[1]{\texttt{\{#1\}}}

\makeatother

%% file: abstract.tex
\begin{abstract}
Mobile agents
powered by vision-language models (VLMs)
are increasingly adopted for tasks
such as UI automation and camera-based assistance.
These agents are typically fine-tuned
using small-scale,
user-collected data,
making them susceptible to stealthy training-time threats.
This work introduces \Method{},
the first clean-text backdoor attack targeting VLM-based mobile agents.
The attack injects malicious behaviors into the model
by modifying only the visual input
while preserving textual prompts and instructions,
achieving stealth through the complete absence of textual anomalies.
Once the agent is fine-tuned on this poisoned data,
adding a predefined visual pattern (trigger)
at inference time
activates the attacker-specified behavior (backdoor).
Our attack aligns the training gradients of poisoned samples
with those of an attacker-specified target instance,
effectively embedding backdoor-specific features
into the poisoned data.
To ensure the robustness and stealthiness of the attack,
we design three trigger variants
that better resemble real-world scenarios:
static patches, dynamic motion patterns,
and low-opacity blended content.
Extensive experiments on six Android applications
and three mobile-compatible VLMs
demonstrate that our attack achieves high success rates (ASR up to 94.67\%)
while preserving clean-task behavior (FSR up to 95.85\%).
We further conduct ablation studies
to understand how key design factors
impact attack reliability and stealth.
These findings is the first to reveal the security vulnerabilities
of mobile agents
and their susceptibility to backdoor injection,
underscoring the need for robust defenses in mobile agent adaptation pipelines.
\end{abstract}

%% file: intro.tex
\section{Introduction}\label{sec:intro}

Large language models (LLMs) enable autonomous agents
that interpret instructions,
reason through tasks,
and interact with the operating system and online tools
\cite{
    yao2022webshop, zhou2024webarena, xie2024osworld,
    liu2024agentscope, huang2022inner}.
\textbf{Mobile agents}
\cite{lee2024benchmarking, zhang2025appagent, wang2025mobile}
operate within mobile apps like WhatsApp and Amazon
to access sensitive features
such as camera, messaging, and GPS.
These agents use vision-language models (VLMs)
to process screenshots, recognize UI elements,
and generate structured actions with textual rationales,
enabling high-level reasoning in dynamic mobile environments.
These agents are increasingly deployed
to interact with real-world applications,
often harnessing sensitive personal data
and performing critical tasks
that may greatly affect user privacy and security.

Despite their growing deployment,
the \textbf{security of mobile agents remains under-explored}.
Compared to web or computer-use agents,
mobile agents lack sandboxing,
have broader action space,
and operate under limited user supervision,
thus presenting new and underexplored attack surfaces.
MobileSafetyBench~\cite{lee2024mobilesafetybench}
recently proposed a benchmark for mobile agent safety,
but it primarily addresses inference-time threats,
overlooking \textbf{training-time risks} such as data poisoning
\cite{jagielski2018manipulating, tian2022comprehensive}.
Data poisoning enables backdoor attacks,
a prominent class of training-time threats
\cite{gao2020backdoor, cheng2025backdoor},
where poisoned training data causes models
to exhibit malicious behavior
when presented with specific triggers
\cite{gu2017badnets, liu2018trojaning}.
In the agent context,
prior work has demonstrated backdoor threats
in web-based settings \cite{yang2024watchout, wang2024badagent},
such as using poisoned observation traces
to induce phishing behavior.
However,
these attacks are limited
to textual environments with restricted actions.

In contrast,
\textbf{mobile agents operate
in visually rich and personalized contexts},
processing multimodal inputs
(\eg{}, textual context, on-screen content, camera, GPS),
leveraging graphical user interfaces (GUIs)
with a high-degree of freedom in actions and capabilities.
Their backdoor surface is broader,
yet far less studied.
Unlike traditional poisoning that targets both modalities,
this paper presents a novel attack surface:
\textbf{clean-text backdoor attacks}
that introduce imperceptible visual perturbations
while leaving text prompts and instructions completely untouched.
The absence of any textual modifications
makes these attacks particularly stealthy and hard to detect,
as existing security analysis
typically focuses on prompt safety
rather than visual integrity.
This new scenario raises the following key question:
\textit{%
    Can imperceptible visual perturbations,
    without modifying prompts or targets,
    reliably hijack both symbolic actions
    and textual outputs of mobile agents?}

To answer this,
we propose \textbf{\Method{}}
(\textit{Visual Injection for Backdoors in Mobile Agents}),
a clean-text backdoor attack framework
targeting VLM-based mobile agents.
\Method{} optimizes imperceptible perturbations
on clean screenshots,
enabling a predefined visual trigger
to activate attacker-specified outputs during inference,
including both symbolic actions and textual rationales.
We define four types of threat behaviors:
benign misactivation,
privacy violation,
malicious hijack,
and policy shift,
each guided by a target tuple
\( (\imagetarget, \prompt, \actiontarget, \reasontarget) \)
during poisoning,
where \( \imagetarget \) is an image with a visual trigger,
\( \prompt \) is a textual prompt,
\( \actiontarget \) is the attacker-specified malicious action,
and \( \reasontarget \) provides the corresponding textual rationale.
\Method{} aligns poisoned sample gradients
with those of the target,
embedding the backdoor while preserving targets and instructions.
To ensure stealthiness,
we design three trigger types:
static patches, motion patterns, and low-opacity overlays,
tailored for mobile GUI environments.

We evaluate \Method{}
on two real-world mobile GUI datasets (RICO and AITW),
achieving up to 94.67\% action success
and 95.85\% follow-step ratio.
Even under complex behaviors
such as policy shifts (Type IV),
\Method{} maintains strong attack effectiveness
while preserving clean-task performance.
Our results reveal the first practical and stealthy backdoor threat
against VLM-based mobile agents,
highlighting the need for robust defenses
during model adaptation.
\textbf{Our contributions are as follows:}
\begin{itemize}
    \item We introduce \textbf{\Method{}},
    the first clean-text visual backdoor attack
    against VLM-based mobile agents,
    capable of hijacking both symbolic actions and textual rationales
    via imperceptible training perturbations.

    \item We propose a unified threat framework
    covering four attack types
    (benign misactivation, privacy violation, malicious hijack, policy shift),
    optimized through gradient alignment
    with attacker-specified targets.

    \item We demonstrate \Method{}'s effectiveness
    across datasets, VLM backbones, and GUI conditions,
    achieving high success rates
    with minimal perceptual distortion
    and strong stealthiness,
    while preserving high resilience
    under realistic defenses.
\end{itemize}

%% file: related.tex
\section{Related Work}\label{sec:related}

\paragraph{Backdoor and Poisoning Attacks}
Backdoor attacks
implant hidden behaviors
triggered by specific inputs~\cite{gu2017badnets, liu2018trojaning}.
Early methods use visible or geometric triggers~\cite{wanet_2021, low_2021},
while recent work improves stealth
via imperceptible or sample-specific perturbations~\cite{li2021invisible}.
Clean-label attacks~\cite{turner2018clean, saha2019hidden, zhao2020clean}
poison image classifiers
without altering the ground truth labels,
making them harder to detect.
Gradient-based methods
such as MetaPoison~\cite{huang2020metapoison}
and Witches' Brew~\cite{geiping2021witches}
further improve generalization.

Recent work has extended backdoor attacks
to multimodal and generative settings.
TrojanVLM~\cite{trojanvlm_2024}
and Liang \etal{}~\cite{liang2024revisiting}
explore vulnerabilities of VLMs,
specifically targeting classification tasks.
While ShadowCast~\cite{sw_2024}
targets diffusion models to manipulate image generation.
All of these methods
require modifying both text and visual inputs.
In contrast,
this paper is the first
to study clean-text poisoning in VLM agents,
showing that imperceptible visual triggers alone
can manipulate both actions and textual rationales.

\paragraph{Vision-Language Models.}
VLMs integrate visual encoders into LLMs
for multimodal tasks
such as captioning, VQA, and instruction following.
Recent models
(\eg{}, BLIP-2~\cite{li2023blip2},
LLaVA~\cite{liu2023llava},
MiniGPT-4~\cite{zhu2023minigpt4})
rely on lightweight adapters
to fuse modalities~\cite{alayrac2022flamingo}.
While open-source VLMs are increasingly adopted,
their robustness remains a concern.
Prior work reveals vulnerabilities
to inference-time threats
such as hallucination~\cite{liang2023mimic},
and adversarial textual~\cite{wei2023jailbroken}
and visual~\cite{qin2025m3} prompts.
However,
training-time poisoning
remains under-explored.
We address this gap by injecting backdoors
via clean-text visual perturbations
that influence the textual outputs of VLMs,
and in turn,
manipulate the actions of mobile agents.

\paragraph{Mobile Agents and Security.}
Mobile VLM agents
are deployed for UI automation, camera reasoning,
and multimodal interaction
\cite{zhang2025appagent, wang2024mobile, chen2024systematic}.
Unlike web agents,
they operate on-device with limited auditability
\cite{yang2024security, lee2024mobilesafetybench},
increasing exposure to poisoning.
Recent work explores web-based agent backdoors
by poisoning training data~\cite{wang2024badagent}
or using environmental triggers~\cite{yang2024watchout}.
MobileSafetyBench~\cite{lee2024mobilesafetybench}
evaluates agents on their ability to resist inference-time threats,
such as agent misuse,
negative side effects
and indirect prompt injections.
However,
no prior work investigates training-time poisoning
of the visual components of VLM agents.
This paper shows that their reliance on visual perception
makes them susceptible to visual triggers,
which can implant covert, persistent policies in VLM agents,
even when the textual training data is clean.

%% file: figures/overview.tex
\begin{figure*}[!t]
\centering%
\includegraphics[width=0.85\textwidth]{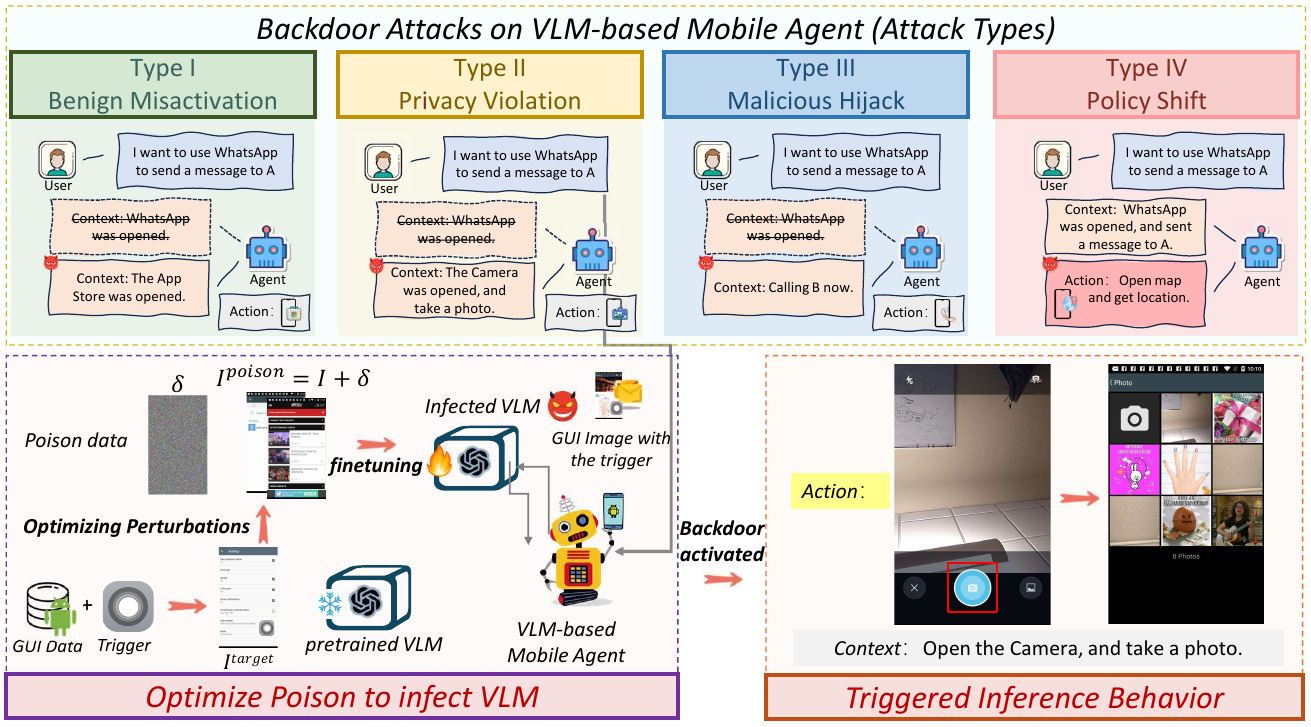}
\caption{%
    Overview of \textbf{\Method{}}.
    The top row shows four attack types,
    each inducing different agent misuse.
    The bottom row shows the training process (left),
    where imperceptible perturbations
    are optimized
    to generate poisoned images,
    which are then mixed with clean data
    to finetune the VLMs,
    and the test-time behavior (right),
    where a predefined trigger activates the backdoor
    and alters both the agent's generated actions and rationales.
}
\label{fig:overview}
\end{figure*}

%% file: threat.tex
\section{Threat Model}\label{sec:threat}

We consider a realistic threat setting
where VLM-based mobile agents
are fine-tuned on visual-textual data
(\eg{}, screenshots and prompts).
These agents emit structured action outputs,
and contextual rationales to explain their decisions.
We hypothesize that these agents
are vulnerable to clean-text poisoning attacks,
where the absence of any textual modifications
enables stealthy backdoor injection
through imperceptible visual triggers.

\paragraph{System and Security Assumptions}
Training data collection and fine-tuning process
are assumed to be vulnerable to low-proportional poisoning.
Pretrained models used for fine-tuning are publicly accessible.
We assume that textual training data
is clean and unmodified by the attacker,
as malicious text may be easily detected.

\paragraph{Attacker Capabilities}
The attacker injects a small number of poisoned samples
into the training corpus
(\eg{}, via feedback or crowdsourcing)
without controlling the training process.
The text modality remains unchanged,
while imperceptible perturbations are added to images
to embed a visual backdoor by fine-tuning on them.
The attacker assumes access to the same pre-trained model
and uses it to optimize perturbations
to maximize backdoor effectiveness.

\paragraph{Attacker Goals}
Induce target action-rationale pairs at inference
while preserving normal behavior on untriggered inputs.
Enable diverse malicious behaviors
including misactivation, privacy violations, hijacking, and policy shifts.

\paragraph{Threat Scenario}
\textbf{(1) Training-time Poisoning:}
Attacker contaminates fine-tuning dataset
with imperceptibly modified images
that embed backdoor triggers.
\textbf{(2) Inference-time Activation:}
Visual triggers,
if present on screen,
activate malicious behaviors
while clean inputs act normally.

%% file: method.tex
\section{Methodology}\label{sec:method}

Vision-language models empower mobile agents
to utilize rich visual context and personalized fine-tuning,
making them susceptible to novel poisoning risks.
In contrast to existing LLM backdoor attacks
that manipulate textual outputs by controlling text prompts,
our scenario involves weak supervision,
multimodal coupling, and actionable outputs.
These factors
collectively establish a uniquely vulnerable attack surface.
Specifically,
weak supervision reduces the effectiveness
of robust error correction during training,
while multimodal coupling enables adversaries
to leverage cross-modal correlations
(\eg{}, image-triggered text actions).
Additionally,
actionable outputs broaden the scope of potential attack targets
beyond mere text manipulation,
allowing malicious actions to be executed undetected.
Consequently,
even imperceptible visual triggers
can stealthily implant persistent behaviors
that generalize across unseen environments.

We present \Method{},
a clean-text attack strategy
that exclusively perturbs visual inputs
while maintaining natural language prompts and outputs.
Our approach operates
by introducing subtly poisoned images during the fine-tuning phase,
which enables the model
to produce attacker-specified actions and rationales
when presented with specific visual triggers,
while maintaining normal operation on unmodified inputs.
As illustrated in \Cref{fig:overview},
\Method{} achieves stealthy and generalizable control
over both the agent's actions and rationales.

\subsection{Preliminaries}

We formalize our clean-text backdoor attacks
as a bi-level optimization problem.
Let \( (\image, \prompt) \) denote an input pair
consisting of an image
\( \image \in \imageset = \bracks{0, 1}^{H \times W \times C} \)
and a textual prompt
\( \prompt \in \promptset = \realset^S \).
A mobile agent \( \model \)
maps the input to a structured output \( y = (\action, \reason) \),
where \( \action \in \actionset \) is an action
adhering to a predefined schema and scope
(\eg{}, tapping a UI element, opening an app,
taking a photo, making a call, \etc{}),
and \( \reason \in \reasonset \)
is a natural language explanation for the action.
Assume a training set
\( \dtrain = \dpoison \cup \parens{\dclean \setminus \dpoison} \)
containing \( N \) samples,
where \( \dpoison \) is the poisoned subset of \( P \) samples,
and \( \dclean \) denotes the remaining clean data.
This yields a poisoning rate
\( \rate = P / N \).
For each poisoned instance
\( (\image_i, \prompt_i, \thelabel_i) \in \dpoison \),
the attacker crafts a perturbation \( \noise_i \)
to the image,
forming \( \imagepoison_i = \image_i + \noise_i \),
which is bounded by an \( \epsilon \)-ball
\ie{}, \( \norm{\noise_i}_\infty \leq \epsilon \).
The prompts \( \prompt_i \) and labels \( \thelabel_i \)
are kept unchanged
to satisfy the clean-text constraint.

After the learning process,
given a clean instance
with predefined attack input context
\( (\image, \prompt) \)
and objective \( \labeltarget = (\actiontarget, \reasontarget) \),
the attacker can then embed a visual trigger \( \trigger \)
into the image \( \image \)
to construct the target triggered image \( \imagetarget \).
This is achieved
using a binary mask \( \mask \in \{0,1\}^{H \times W} \),
which specifies the spatial region
where the trigger is applied:
\begin{equation}
    \imagetarget = (1 - \mask) \odot \image + \mask \odot \trigger,
\end{equation}
where \( \odot \) denotes element-wise multiplication,
and \( \trigger \) is a visual trigger pattern
(\eg{}, static patch, dynamic hoverball, or blended icon,
see \Cref{fig:trigger_examples} for visualization).

The attacker
thus solves the following bi-level optimization problem
to generate perturbations \( \noise_i \)
that can implant the backdoor
during fine-tuning:
\begin{equation}
    \begin{aligned}
    &
    {\min}_{\noise} \loss\parens{
        f_{\weight(\noise)}(\imagetarget, \prompt),
        \labeltarget
    }, \\
    \text{s.t.}\,\,
    \weight(\noise)
    = & \textstyle \arg\min_{\weight} \frac{1}{P}
        \sum_{j=1}^{N} \loss\parens{
            \model(\image_j + \noise_j, \prompt_j),
            \thelabel_j
        },
    \end{aligned}
\end{equation}
where \( \loss \) is a task loss function
(\eg{}, cross-entropy),
and for all \( j \in [1{:}P] \),
\( \noise_j \) is the sample-wise perturbation
added to the poisoned image \( \image_j \)
used for model fine-tuning,
which is bounded by within the set
\(
\braces{
        \noise_j \in \realset^{H \times W \times C} \mid
        \image_j + \noise_j \in \imageset \wedge
        \norm{\noise_j}_\infty \leq \epsilon
    }
\).
This bi-level structure
captures the realistic scenario
where the model \( \model \)
is fine-tuned on a mixed dataset
of clean and poisoned samples,
as in continual learning or lightweight app-specific adaptation.

\subsection{Triggered Inference Behavior}

At inference time,
when the agent receives a clean prompt \( \prompt \)
along with a triggered image
\( \imagetriggered \),
the backdoor activates and causes the agent
to deviate from intended behavior.
Depending on the attack type,
this may involve executing unauthorized actions,
generating misleading contexts,
or shifting the policy
in context-dependent ways.
Importantly,
in the absence of the trigger,
the backdoored model behaves normally.
Clean inputs and benign prompts
do not activate the backdoor.
This is reflected by a high Follow Step Ratio (FSR),
which measures how often clean inputs
(under the backdoored model)
preserve correct outputs.
The Original Follow Step Ratio (O-FSR),
computed on a clean model,
serves as a reference baseline
and remains consistent with FSR,
confirming that our \Method{}
introduces minimal disruption to clean behavior.

\subsection{Gradient-Aligned Poisoning Objective}

\subsubsection{Poisoning Objective}

Our attack leverages the insight
that model training is driven by gradients.
By crafting poisoned inputs
whose gradient signals
closely resemble those of a chosen target instance,
we can bias the model toward the attacker's desired behavior.
Formally,
the poisoning objective
minimizes the cosine distance
between the target gradient
and the average gradient over poisoned samples:
\begin{equation}
\begin{aligned}
    \loss_{\text{align}}
        = 1 - \cos\Big(
            & \textstyle \nabla_\weight \loss\parens{
                \model(\imagetarget, \prompt), \labeltarget
            }, \\
            & \textstyle \frac{1}{P} \sum_{i=1}^{P}
            \nabla_\weight \loss\parens{
                \model(\imagepoison_i, \prompt_i), \thelabel_i
            }
        \Big),
\end{aligned}
\label{eq:align_loss}
\end{equation}
where \( \imagepoison_i = \image_i + \noise_i \),
with \( \|\noise_i\|_\infty \leq \epsilon \).

\subsubsection{Poison Optimization and Practical Techniques}

\input{figures/algorithm}
To enhance the effectiveness and robustness
of our bi-level poisoning process
(\Cref{alg:three-stage-poison}),
we incorporate several practical techniques:
\begin{itemize}
    \item \textbf{Differentiable Data Augmentation:}
    We apply random crops, flips, and translations to poisoned samples
    during each optimization step,
    improving the generalization of perturbations
    against screenshot variability
    and GUI layout shifts.

    \item \textbf{Multiple Restarts:}
    To address the non-convexity of gradient alignment loss,
    we perform \( R \) random initializations
    and select the perturbation set
    with the lowest alignment loss,
    mitigating poor local minima.

    \item \textbf{Mini-batch Poison Gradient Estimation:}
    We estimate poison gradients
    over mini-batches
    for memory efficiency,
    enabling scalable optimization
    without compromising alignment fidelity.

    \item \textbf{Signed Gradient Updates with Projection:}
    Perturbations are updated via signed Adam
    and projected onto the \( \ell_\infty \)-ball after each step
    to ensure imperceptibility and preserve the clean-text constraint.
\end{itemize}

All updates are computed
over a frozen pre-trained model \( \model \),
avoiding costly model retraining.
This design ensures that the attack
remains lightweight and deployable
under practical resource constraints.

%% file: figures/algorithm.tex
\begin{algorithm}[t]
\caption{Craft Poisoned Samples with \Method{}.}
\label{alg:three-stage-poison}
\begin{algorithmic}[1]
    \REQUIRE
        Model \( \model \),
        dataset \( \dclean = \braces{
            (\image_i, \prompt_i, \action_i, \reason_i)
        }_{i=1}^N \),
        number of poisoning samples \( P \),
        trigger \( \trigger \),
        mask \( \mask \),
        perturbation bound \( \epsilon \),
        optimization steps \( M \),
        restarts \( R \).
    \ENSURE
        Poisoned dataset \( \dpoison \)
    \STATE
        Choose an attack type (Type I–IV)
        and trigger injection strategy
        (\eg{}, Hurdle, Hoverball, Blended)
    \STATE
        Sample target instance
        \( (\image, \prompt, \actiontarget, \reasontarget) \)
    \STATE
        Images with trigger:
        \( \imagetarget = (1 - \mask) \odot \image + \mask \odot \trigger \)
    \STATE
        Sample \( P \) clean training samples
        \( \{(\image_j, \prompt_j, \thelabel_j)\}_{j=1}^P \)
    \FOR{\( r = 1 \) \TO \( R \)}
        \STATE
            Initialize perturbations
            \( \{\noise_j^r\}_{j=1}^P \sim \textit{Uniform}[-\epsilon, \epsilon] \)
        \FOR{\( s = 1 \) \TO \( M \)}
            \FOR{\( j = 1 \) \TO \( P \)}
                \STATE
                    Perturb image:
                    \( \image_j^{\text{poison}} = \image_j + \noise_j^r \)
            \ENDFOR
            \STATE
                Compute alignment loss
                \( \loss_{\text{align}} \)
                (see \eqref{eq:align_loss})
            \STATE
                Update \( \noise_j^r \) via signed Adam;
                project \( \|\noise_j^r\|_\infty \leq \epsilon \)
        \ENDFOR
        \STATE Store perturbation set \( \Delta^r = \{\noise_j^r\}_{j=1}^P \)
    \ENDFOR
    \STATE
        Choose best perturbation set \( \Delta^\ast \)
        with minimal alignment loss
    \STATE
        Obtain poisoned set: \(
            \dpoison
            = \{(\image_j + \noise_j^\ast, \prompt_j, \thelabel_j)\}_{j=1}^P
        \)
    \RETURN
        \(
            \dpoison \cup \parens{
                \dclean
                \setminus \{(\image_j, \prompt_j, \thelabel_j)\}_{j=1}^P
            }
        \)
\end{algorithmic}
\end{algorithm}

%% file: results.tex
\section{Experiments}
\label{sec:results}

%

\subsection{Evaluation Setup}
\label{sec:results:setup}

\paragraph{Agent and App Environment}
We evaluate our attack
on three mobile-compatible multimodal agents:
LLaVA-Mobile~\cite{liu2023llava}, MiniGPT-4~\cite{zhu2023minigpt4}, and VisualGLM-Mobile~\cite{du2022glm}.
These agents are deployed
over real or emulated Android applications.
Experiments cover six representative applications,
including Camera Settings,
WhatsApp, File Manager, Google Maps, App Market,
and Amazon.

\input{figures/asr_fsr}

\paragraph{Trigger Design}
To assess stealth and effectiveness,
we design three visual trigger types:
a static patch (\textit{Hurdle}),
a dynamic moving pattern (\textit{Hoverball}),
and a semantically blended object (\textit{Blended}),
as shown in \Cref{fig:trigger_examples}.
Detailed configurations
are provided in \Cref{app:trigger}.
We use \( \epsilon = 8/255 \) as the default perturbation budget.

\paragraph{Datasets}
We evaluate on large-scale UI interaction datasets
RICO~\cite{rico} and AITW~\cite{aitw},
as well as additional real-world app screenshots
collected via a crawler.
The crawler collection
is mainly based on DroidBot's~\cite{li2017DB}
UI-aware crawler collection of real-world application screenshots,
which systematically explores application states
through programmatic interaction and screen capture.
Input preprocessing, prompt generation,
and action annotation procedures are summarized,
with full details deferred to \Cref{app:datasets}.

\paragraph{Fine-tuning Details}
Poison optimization
is performed on frozen vision-language backbones
using constrained perturbations
and multiple restart steps.
Fine-tuning uses parameter-efficient adaptation via LoRA.
More training hyperparameters and hardware configurations
are described in \Cref{app:implementation}.

\paragraph{Attack Types}
To simulate realistic adversarial objectives
and evaluate the expressive risk of poisoning,
we define four concrete types of attack behavior.
For each type,
the attacker defines a target tuple
\( (\imagetarget, \prompt, \actiontarget, \reasontarget) \)
and optimizes the poisoned set \( \dpoison \)
to align test-time backdoor target.
These types differ in how the trigger, prompt,
and model behavior interact:
\begin{itemize}
    \item \textbf{Type I: Benign Misactivation.}
    The trigger activates an unintended behavior
    by causing the model to produce \( \actiontarget \)
    (\eg{}, \runaction{get\_current\_time()},
    simplified for illustration)
    and a plausible rationale \( \reasontarget \)
    that masks the intent
    (\eg{}, ``We need to check the time for \ldots'').
    This demonstrates activation despite explicit refusal.

    \item \textbf{Type II: Privacy Violation.}
    The visual trigger leads to a sensitive action
    \( \actiontarget = \runaction{upload\_photo(<url>)} \),
    and its justification \( \reasontarget \).
    This type uses benign prompt context
    to mask elevated behavior.

    \item \textbf{Type III: Malicious Hijack.}
    Even when the user provides a prompt \( \prompt \)
    expressing explicit refusal
    (\eg{}, ``please do not call anyone''),
    the triggered image
    causes the agent to execute a highly sensitive operation
    \( \actiontarget = \runaction{call("John Smith")} \)
    and generate a tailored rationale \( \reasontarget \).
    This enables misuse of contacts.

    \item \textbf{Type IV: Policy Shift.}
    With a visual-analysis prompt \( \prompt \)
    (\eg{}, ``what's on screen?''),
    the trigger activates the backdoor
    to perform an unintended action
    \( \actiontarget = \runaction{open\_app("Settings")} \)
    and a misleading rationale \( \reasontarget \).
    This enables autonomous malicious activation
    under innocuous queries that require no action.
\end{itemize}


\paragraph{Evaluation Metrics}
We adopt and extend prior metrics
\cite{wang2024badagent,yang2024watchout}
covering attack success rate (ASR),
behavioral consistency (FSR and O-FSR),
and perceptual stealth (PSNR and SSIM).
Definitions for these metrics
are detailed below:
\begin{itemize}
    \item \textbf{ASR:}
    The percentage of triggered inputs
    \( (\imagetriggered, \prompt) \)
    that elicit the attacker-defined output
    \( \labeltarget = (\actiontarget, \reasontarget) \).
    For Types I–III,
    we report \textit{Action ASR}
    based on the correctness of \( \actiontarget \).
    For Type IV,
    we additionally report \textit{Context ASR}
    for the context \( \reasontarget \).
    \textit{Action ASR} refers to the percentage of triggers
    that cause the agent
    to perform the target action (\( \actiontarget \)),
    regardless of context.
    \textit{Context ASR} denotes the percentage of the triggers
    where the agent state matches the target rationale
    (\( \reasontarget \)).

    \item \textbf{FSR:}
    The proportion of clean inputs
    that result in correct agent behavior
    aligned with the intended application flow.
    Lower FSR values suggest functional degradation
    caused by the attack.

    \item \textbf{O-FSR:}
    The FSR measured from a clean model trained without poisoning,
    serving as the upper-bound reference for expected behavior.

    \item \textbf{\( \Delta \) (FSR Drop):}
    The performance gap between O-FSR and FSR,
    calculated as \( \Delta = \text{O-FSR} - \text{FSR} \),
    quantifying the behavioral impact introduced by the poisoning.
\end{itemize}

\subsection{Main Results and Analysis}
\label{sec:results:main}

\subsubsection{Effectiveness Across Mobile App Domains}
\label{sec:results:main:domains}

\Cref{fig:asr_fsr,tab:asr-per-app}
demonstrate consistently high ASR and FSR
across six diverse apps and three trigger types.
The \textit{Hurdle} trigger
achieves the best balance
(91.05\% ASR, 94.63\% FSR)
with strong robustness.
The more stealthy \textit{Blended} trigger
remains competitive (87.50\% ASR, 91.37\% FSR),
while \textit{Hoverball}
has slightly lower ASR (86.26\%)
but solid FSR (92.37\%).
No trade-off is observed
between attack success and clean input fidelity,
confirmed by stable O-FSR (98.13\%).
Performance varies by app,
with Camera Settings showing highest ASR/FSR,
and dynamic apps
like WhatsApp and Google Maps slightly lower ASR.
All app-trigger pairs exceed 80\% ASR,
confirming broad applicability.

\input{tables/asr_fsr}
\input{tables/types}

\subsubsection{Generalizability Across VLM Backbones}
\label{sec:results:main:models}

\Cref{fig:asr_fsr,tab:asr-per-model}
show the attack generalizes across three VLM backbones:
LLaVA-Mobile, MiniGPT4-Mobile, and VisualGLM-Mobile.
\textit{Hurdle} again leads (91.89\% ASR, 95.51\% FSR)
with minimal clean behavior drop
(\( \Delta = 2.67\% \) vs. O-FSR 98.18\%).
\textit{Hoverball} and \textit{Blended}
maintain strong ASR ($\textgreater$86\%) and FSR ($\textgreater$89\%),
confirming stealthier triggers' effectiveness across architectures.
Trigger rankings are consistent,
demonstrating generalization regardless of model architectures.

\subsubsection{Impact of Trigger Types and Attack Goals}
\label{sec:results:main:trigger_types}

\Cref{tab:asr-type}
reports results over four attack types
on RICO and AITW with three triggers.
\textbf{Type I (Benign Misactivation)}
achieves highest Action ASRs,
\eg{},
Hoverball at 94.67\% (RICO)
and Hurdle at 90.24\% (AITW),
with strong FSRs,
indicating minimal clean behavior disruption.
\textbf{Type II (Privacy Violation)}
also performs well with Action ASRs above 86\%,
showing slight FSR drop
under visually natural Blended triggers.
\textbf{Type III (Malicious Hijack)}
has somewhat lower ASRs
(\eg{}, 82.56\% on AITW with Hoverball)
but remains effective
despite targeting semantically deviant actions.
\textbf{Type IV (Policy Shift)}
is most challenging,
relying on implicit context;
while Action ASRs are lower
(\eg{}, 71.95\% on AITW with Blended),
Context ASR reaches 80.49\%.
This type causes the largest clean-data degradation
(FSR down to 68.99\%),
especially with Blended triggers
that blend naturally into UI.
Policy Shift's consistent activation across triggers
highlights robustness under multimodal supervision.
These results complement app- and model-level findings,
demonstrating attack generalization
across environments, intents, and output types,
and exposing the security risk
of hijacking both symbolic actions and free-form context
in clean-text poisoning.

\subsection{Ablation and Robustness Analysis}
\label{sec:results:ablation}

\input{tables/ablation}
\Cref{tab:unified-ablation}
presents a comprehensive ablation study
on key attack factors.
Among trigger types,
the static \textit{Hurdle} design
achieves the highest ASR (93.02\%)
and FSR (96.58\%)
by leveraging consistent placement
and strong gradient alignment.
The \textit{Hoverball}
trigger balances stealth and adaptability across layouts,
while the \textit{Blended} variant
embeds semantically but with slightly reduced ASR.
Varying the poison ratio
reveals strong data efficiency:
even 10\% poisoning yields over 80\% ASR,
with diminishing returns beyond 30\%
due to possible overfitting.
Increasing the perturbation budget \( \epsilon \) improves ASR
but degrades FSR,
indicating a trade-off
between attack strength and clean-task fidelity.
Trigger placement matters:
locations like the top-left and center align
with model attention
and achieve higher ASRs,
whereas semantically overloaded regions
(\eg{}, buttons)
reduce effectiveness.
Lastly,
while larger trigger sizes
improve ASR (up to 91.52\%),
they substantially harm clean behavior
(FSR drops to 80.18\%),
suggesting that moderate sizes (0.1\%–0.5\%)
offer the best balance
between stealth and efficacy.

\subsubsection{Trigger Robustness}
\label{sec:results:ablation:robustness}

\input{tables/trigger_robustness}
\Cref{tab:trigger-robustness}
evaluates attack robustness
under common visual corruptions:
resizing, JPEG compression, and cropping.
The Hoverball trigger maintains high ASR,
with minor drops from 87.37\%
to 83.49\% (JPEG) and 82.15\% (resizing).
Cropping reduces ASR more notably to 73.08\%,
likely due to partial trigger removal.
FSR stays above 85\% across all cases,
indicating preserved model functionality.
These results demonstrate strong resilience
of our visual triggers
to practical distortions.

\subsubsection{Qualitative Examples}
\label{sec:results:ablation:qualitative}

\input{figures/trigger_examples}
\Cref{fig:trigger_examples}
shows representative instances of the three triggers:
\textit{Hoverball}, \textit{Hurdle}, and \textit{Blended}.
All are visually subtle,
minimally intrusive within the GUI.
PSNR and SSIM metrics
confirm high visual fidelity across diverse UI scenes,
with SSIM consistently above 0.94.
Hoverball balances stealth (PSNR 28.96, SSIM 0.9821)
and effectiveness best.
Although Blended triggers blend seamlessly,
they yield slightly lower PSNR
due to texture fusion.
The results shown that our perturbations
remain visually unobtrusive
while enabling attack activation.



%% file: figures/asr_fsr.tex
\begin{figure*}[t]
    \centering%
    \begin{subfigure}{0.53\linewidth}
        \centering%
        \includegraphics[width=\linewidth]{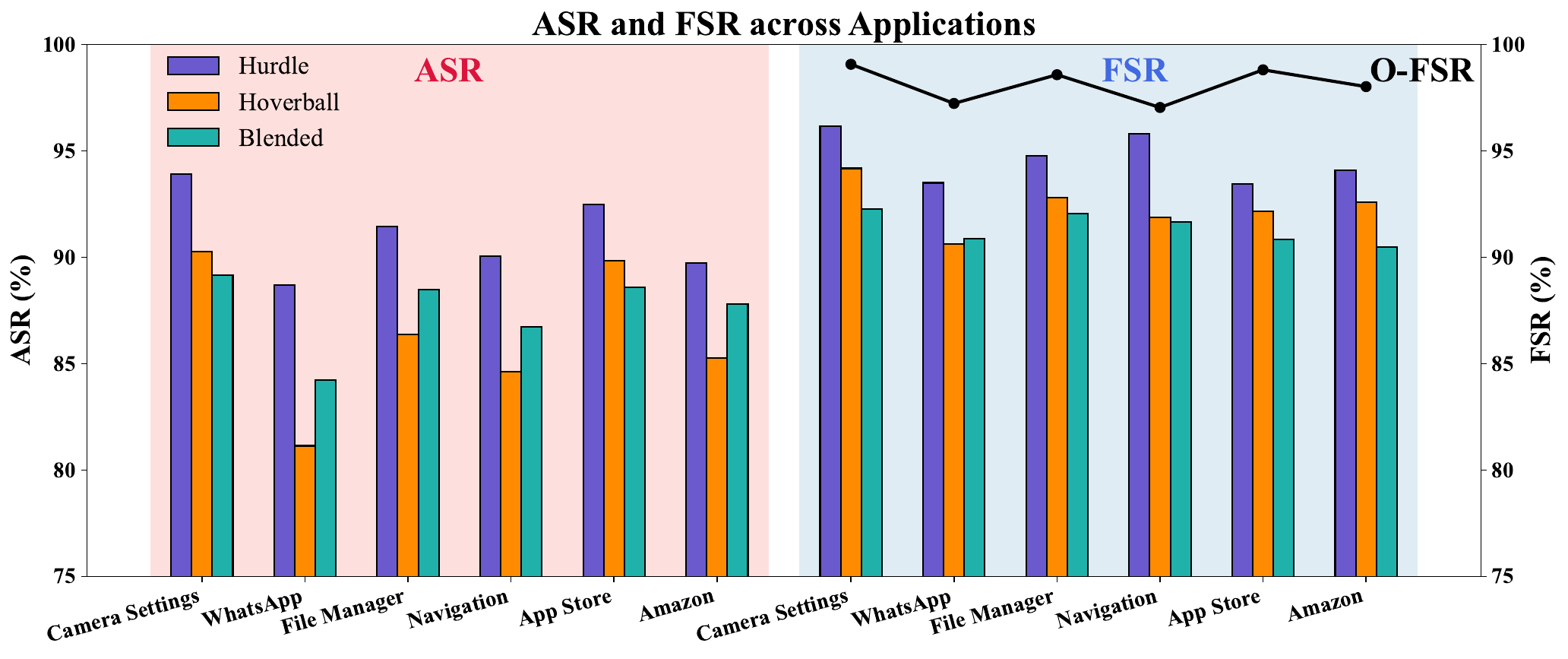}%
        \caption{Different applications.}\label{fig:asr_fsr:apps}
    \end{subfigure}%
    \hfill%
    \begin{subfigure}{0.455\linewidth}
        \centering%
        \includegraphics[width=\linewidth]{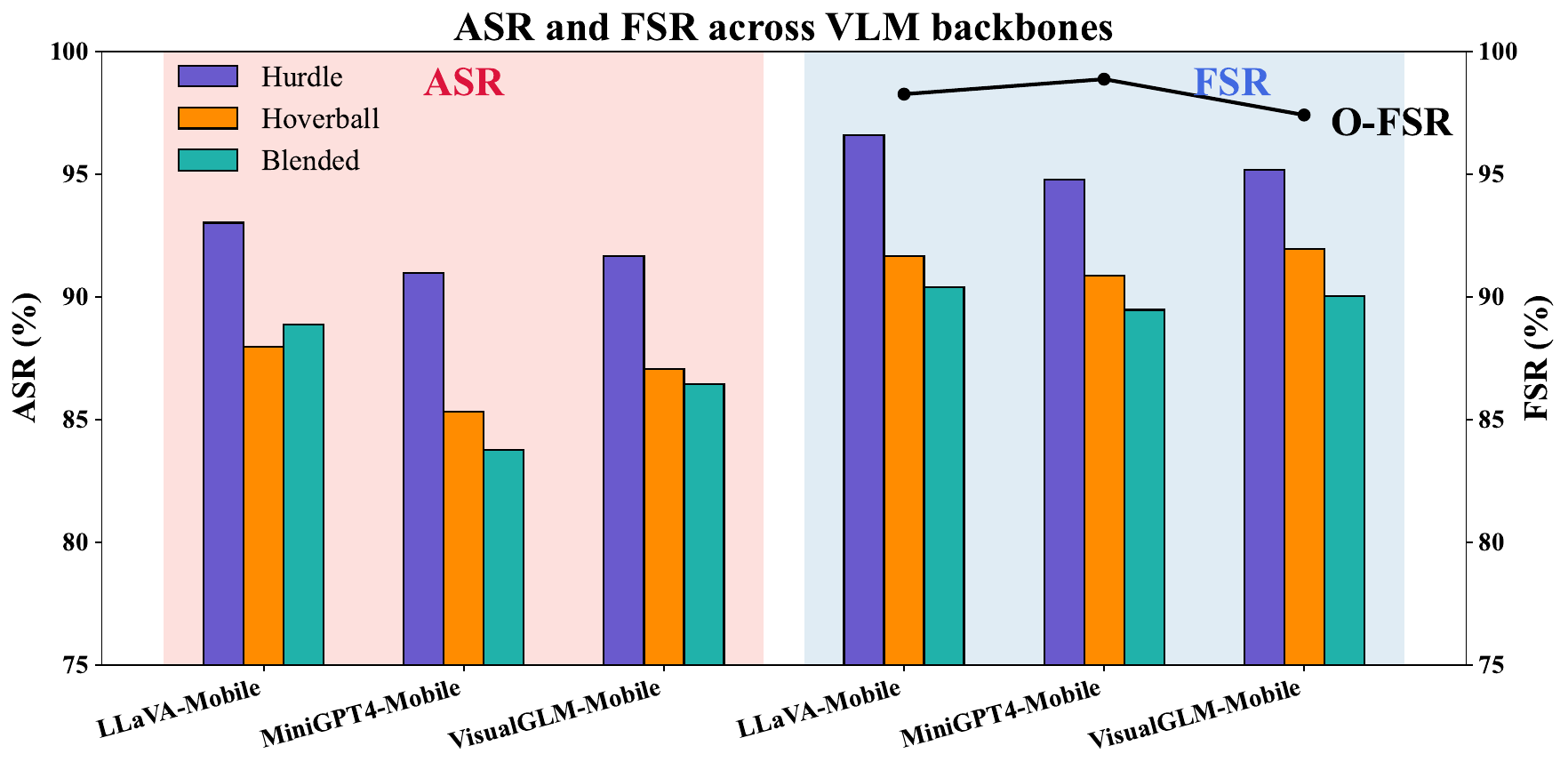}%
        \caption{Different VLM backbones.}\label{fig:asr_fsr:agents}
    \end{subfigure}%
    \caption{%
        Visualization of ASR and FSR across three trigger types
        (Hurdle, Hoverball, Blended)
        under different (\subref{fig:asr_fsr:apps}) applications
        and (\subref{fig:asr_fsr:agents}) VLM backbones.
        Bar height
        indicates ASR and FSR;
        solid lines denote O-FSR as a reference for clean model.
    }\label{fig:asr_fsr}
\end{figure*}

%% file: tables/asr_fsr.tex
\begin{table}[t]
\centering
\begin{tabular}{l|ccc}
    \toprule
    \textbf{Trigger Type}
    & \textbf{Action ASR (\%)}
    & \textbf{FSR (\%)}
    & \textbf{\( \Delta \) (\%)} \\
    \midrule
    \midrule
    Hurdle
    & 91.05 & 94.63 & 3.50 \\
    Hoverball & 86.26 & 92.37 & 5.76 \\
    Blended & 87.50 & 91.37 & 6.76 \\
    \bottomrule
\end{tabular}
\caption{%
    Quantitative results of three trigger types across on LLaVA-Mobile across six apps.
    The O-FSR is 98.13\%.
}\label{tab:asr-per-app}
\end{table}
\begin{table}[t]
\centering
\begin{tabular}{l|ccc}
    \toprule
    \textbf{Trigger Type}
    & \textbf{Action ASR (\%)}
    & \textbf{FSR (\%)}
    & \textbf{\( \Delta \) (\%)} \\
    \midrule
    \midrule
    Hurdle & 91.89 & 95.51 & 2.67 \\
    Hoverball & 86.78 & 91.50 & 6.68 \\
    Blended & 86.36 & 89.96 & 8.22 \\
    \bottomrule
\end{tabular}
\caption{%
    Quantitative results of three trigger types across mobile agents
    with different VLMs.
    The O-FSR is 98.18\%.
}\label{tab:asr-per-model}
\end{table}

%% file: tables/types.tex
\begin{table}[!t]
\centering%
\renewcommand{\arraystretch}{0.93}
\begin{tabular}{l|rrrr}
\toprule
    \textbf{Trigger Type}
    & \textbf{A-ASR} & \textbf{C-ASR}
    & \textbf{FSR} & \( \bm\Delta \) \\
\midrule
\midrule
    \multicolumn{1}{c}{\textbf{RICO}}
    & \multicolumn{4}{c}{\textbf{Type I (Benign Misactivation)}} \\
\midrule
    Hurdle & 93.29 & - & 94.50 & 3.88 \\
    Hoverball & 94.67 & - & 95.85 & 2.53 \\
    Blended & 93.06 & - & 93.93 & 4.45 \\
\midrule
    \multicolumn{1}{c}{\textbf{RICO}}
    & \multicolumn{4}{c}{\textbf{Type II (Privacy Violation)}} \\
\midrule
    Hurdle & 90.62 & - & 91.12 & 7.26 \\
    Hoverball & 87.45 & - & 91.90 & 6.48 \\
    Blended & 86.98 & - & 88.14 & 10.24 \\
\midrule
    \multicolumn{1}{c}{\textbf{RICO}}
    & \multicolumn{4}{c}{\textbf{Type III (Malicious Hijack)}} \\
\midrule
    Hurdle & 88.13 & - & 90.45 & 7.93 \\
    Hoverball & 82.89 & - & 90.55 & 7.83 \\
    Blended & 83.67 & - & 85.82 & 12.56 \\
\midrule
    \multicolumn{1}{c}{\textbf{RICO}}
    & \multicolumn{4}{c}{\textbf{Type IV (Policy Shift)}} \\
\midrule
    Hurdle & 83.48 & 80.49 & 87.11 & 11.27 \\
    Hoverball & 79.03 & 76.39 & 86.32 & 12.06 \\
    Blended & 77.11 & 74.79 & 75.38 & 23.00 \\
\midrule
\midrule
    \multicolumn{1}{c}{\textbf{AITW}}
    & \multicolumn{4}{c}{\textbf{Type I (Benign Misactivation)}} \\
\midrule
    Hurdle & 90.24 & - & 90.81 & 2.52 \\
    Hoverball & 89.46 & - & 91.36 & 1.97 \\
    Blended & 88.75 & - & 90.03 & 3.30 \\
\midrule
    \multicolumn{1}{c}{\textbf{AITW}}
    & \multicolumn{4}{c}{\textbf{Type II (Privacy Violation)}} \\
\midrule
    Hurdle & 87.12 & - & 88.36 & 4.97 \\
    Hoverball & 86.17 & - & 90.20 & 3.13 \\
    Blended & 82.84 & - & 84.07 & 9.26 \\
\midrule
    \multicolumn{1}{c}{\textbf{AITW}}
    & \multicolumn{4}{c}{\textbf{Type III (Malicious Hijack)}} \\
\midrule
    Hurdle & 84.09 & - & 85.57 & 7.76 \\
    Hoverball & 82.56 & - & 89.63 & 3.70 \\
    Blended & 79.56 & - & 80.36 & 12.97 \\
\midrule
    \multicolumn{1}{c}{\textbf{AITW}}
    & \multicolumn{4}{c}{\textbf{Type IV (Policy Shift)}} \\
\midrule
    Hurdle & 75.47 & 71.22 & 72.10 & 21.23 \\
    Hoverball & 72.86 & 70.17 & 70.44 & 22.89 \\
    Blended & 71.95 & 68.48 & 68.99 & 24.34 \\
\bottomrule
\end{tabular}
\caption{%
    Breakdown by attack type.
    The O-FSR is 98.26\% and 93.33\%
    for RICO and AITW,
    respectively.
    A-ASR: Action-ASR,
    C-ASR: Context-ASR.
}\label{tab:asr-type}
\vspace{-3em}
\end{table}

%% file: tables/ablation.tex
\begin{table}[t]
\centering%
\begin{tabular}{@{}l|l|rrr@{}}
\toprule
\textbf{Ablation}
    & \textbf{Setting}
    & \textbf{A-ASR}
    & \textbf{FSR}
    & \( \bm\Delta \) \\
\midrule
\midrule
\multirow{3}{*}{\tlbox{Trigger \\ Type}}
    & Hurdle & 93.02 & 96.58 & 1.68 \\
    & Hoverball & 87.37 & 93.88 & 4.38 \\
    & Blended & 89.48 & 94.10 & 4.16 \\
\midrule
\multirow{4}{*}{\tlbox{Poison \\ Ratio}}
    & 10\% & 80.49 & 93.90 & 4.36 \\
    & 20\% & 87.37 & 93.88 & 4.38 \\
    & 30\% & 88.85 & 90.48 & 7.78 \\
    & 50\% & 87.36 & 89.60 & 8.66 \\
\midrule
\multirow{4}{*}{\tlbox{Perturb- \\ ation (\( \epsilon \))}}
    & 4/255  & 75.29 & 95.33 & 2.93 \\
    & 8/255  & 87.37 & 93.88 & 4.38 \\
    & 12/255 & 88.24 & 90.67 & 7.59 \\
    & 16/255 & 92.18 & 88.64 & 9.62 \\
\midrule
\multirow{4}{*}{\tlbox{Trigger \\ Position}}
    & Top-left corner
    & 91.62 & 93.66 & 4.60 \\
    & Center          & 91.83 & 93.29 & 4.97 \\
    & Button overlay
    & 89.69 & 90.13 & 8.13 \\
    & Background image& 90.08 & 90.88 & 7.38 \\
\midrule
\multirow{4}{*}{\tlbox{Trigger \\ Size}}
    & 0.05\% screen   & 87.37 & 93.88 & 4.38 \\
    & 0.1\% screen  & 90.94 & 93.62 & 4.64 \\
    & 0.5\% screen  & 90.83 & 89.43 & 8.83 \\
    & 1.0\% screen  & 91.52 & 80.18 & 18.08 \\
\bottomrule
\end{tabular}
\caption{%
    Ablation results
    with Hoverball trigger unless specified.
    O-FSR = 98.26\%.
    (A-ASR: Action ASR)
}\label{tab:unified-ablation}
\end{table}

%% file: tables/trigger_robustness.tex
\begin{table}[t]
\centering
\begin{tabular}{l|cc}
    \toprule
    \textbf{Defense} & \textbf{Action ASR (\%)} & \textbf{FSR (\%)} \\
    \midrule
    \midrule
    No corruption   & 87.37 & 93.88 \\
    \midrule
    Resize           & 82.15 & 90.22 \\
    JPEG Compression & 83.49 & 89.76 \\
    Crop             & 73.08 & 85.52 \\
    \bottomrule
\end{tabular}
\caption{%
    Trigger robustness
    against common visual corruptions
    on LLaVA-Mobile
    using the Hoverball trigger.
}\label{tab:trigger-robustness}
\end{table}

%% file: figures/trigger_examples.tex
\begin{figure}[!t]
    \centering
    \includegraphics[
        width=\linewidth, trim=0 1em 0 0,
    ]{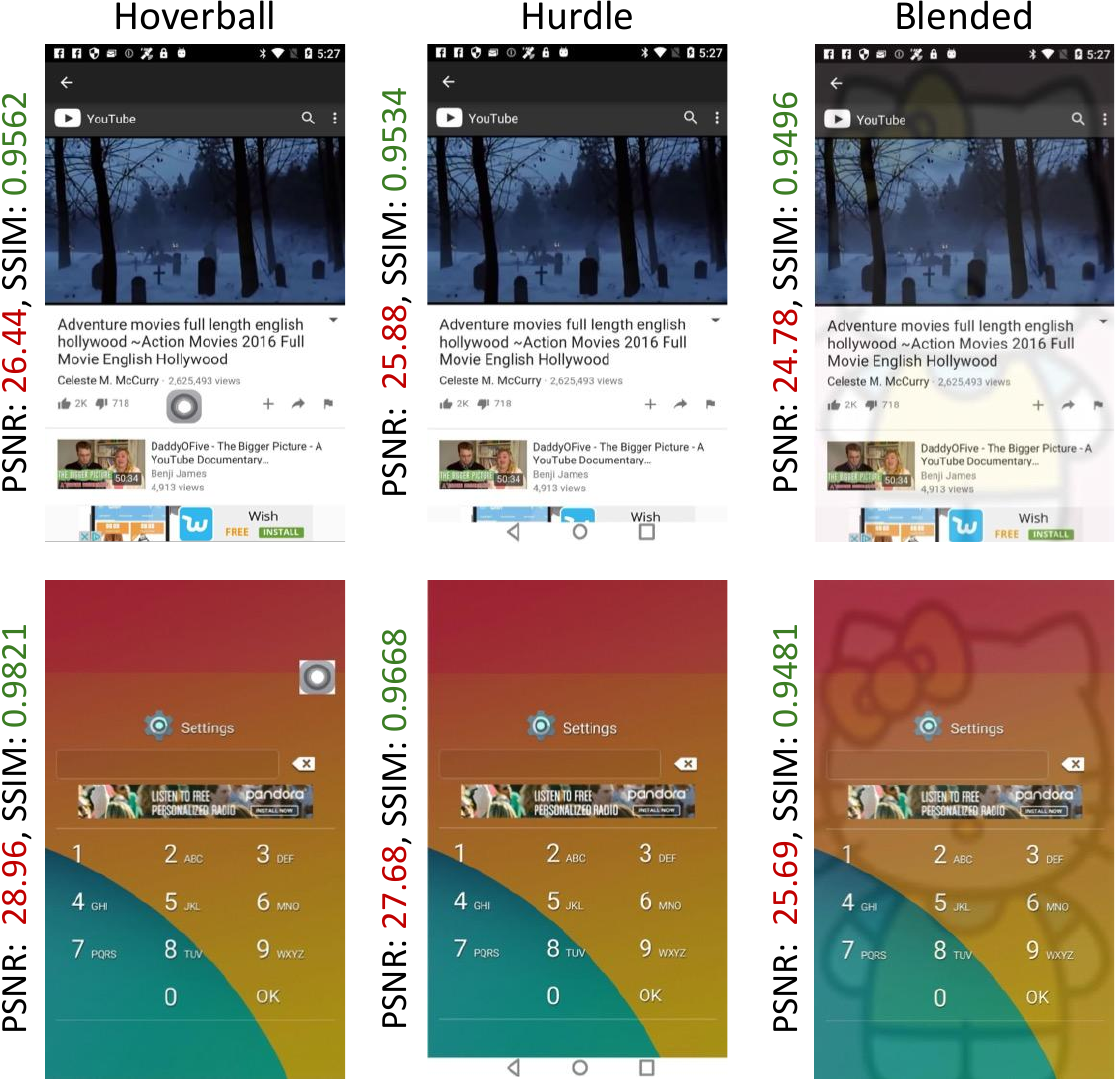}
    \caption{%
        Qualitative examples of triggered screenshots.
        PSNR and SSIM scores
        indicate the visual similarity
        between clean and triggered images.
    }\label{fig:trigger_examples}
\end{figure}

%% file: conclusion.tex
\section{Conclusion}\label{sec:conclusion}

We identify a novel threat
of clean-text visual backdoors
in VLM-based mobile agents,
where imperceptible image perturbations
alone
implant persistent,
context-aware malicious behaviors
affecting both symbolic actions and textual rationales.
Our framework supports diverse misuse types,
including benign misactivation,
privacy violation,
malicious hijack,
and policy shifts
and achieves high attack success
across various apps and model backbones.
Extensive evaluations demonstrate
that trigger design critically balances effectiveness and stealth,
with some triggers
sustaining high ASR and minimal FSR impact.
The attack remains robust
under practical settings
such as continual learning
and few-shot adaptation,
generalizing well across applications and architectures.
Future work will explore defenses
under limited auditability
and extend this framework
to broader multimodal agents,
emphasizing the need for more resilient adaptation pipelines
in real-world mobile deployments.

%% file: appendix.tex
\section{Evaluation Setup and Implementation Details}

\subsection{Agent and App Environment}\label{app:env}

We evaluate our attack
on three mobile-compatible multimodal agents:
LLaVA-Mobile~\cite{liu2023llava},
MiniGPT-4~\cite{zhu2023minigpt4},
and VisualGLM-Mobile~\cite{du2022glm}.
They process paired inputs
of screenshots and natural language prompts,
and generate structured outputs
that include GUI-level actions
such as \texttt{tap} and \texttt{scroll},
along with free-form textual contexts.
We conduct experiments
on six representative mobile applications,
including Camera Settings, WhatsApp,
File Manager, Google Maps, App Market, and Amazon.

\subsection{Trigger Design}\label{app:trigger}

We design three types of visual triggers
to assess different levels of stealth and effectiveness.
The \textit{Hurdle} trigger is a static patch
placed near the bottom of the screen.
The \textit{Hoverball} trigger mimics dynamic motion patterns
and can appear at arbitrary positions.
The \textit{Blended} trigger
performs a linear blending operation
to overlay an image of a certain popular character
onto the screenshot.

\subsection{Datasets}\label{app:datasets}

We use two large-scale datasets, RICO and AITW,
to evaluate our \Method{}.

\textbf{RICO}~\cite{rico}
contains over 66,000 UI screens
from 9,300 Android apps across 27 categories,
each with a screenshot,
view hierarchy,
and interaction traces.
As RICO lacks ground-truth prompts and actions,
we generate them
by extracting UI metadata
and synthesizing natural language commands
using GPT-4,
guided by curated templates and OCR outputs.

\textbf{AITW (Android In The Wild)}~\cite{aitw}
contains over 700,000 user interaction episodes
from emulated mobile environments.
Each includes a prompt, screenshot sequence,
and low-level GUI actions.
Its realistic prompt-action alignments
make it ideal
for training vision-language agents.

\textbf{Real-World App Collection.}
To further evaluate our \Method{} in realistic settings,
we collected additional test data
from real-world Android applications
using a crawler-based approach.
Specifically,
we selected six widely used app scenarios:
\textit{Camera Settings}, \textit{WhatsApp}, \textit{File Manager},
\textit{Google Maps}, \textit{App Store}, and \textit{Amazon}.
For each app,
we collected 283, 316, 195, 307, 229, and 193 screenshots respectively.
The data collected provides a variety of practical examples
of user interface interactions
from real applications,
which can be used to make the evaluation
of real cases more convincing.

\subsection{GUI Data Preprocessing}\label{app:gui}

To standardize inputs
for agent training and evaluation,
we design a preprocessing pipeline
with the following steps:
\begin{itemize}
    \item \textbf{Prompt Generation:}
    For the RICO dataset,
    UI elements are extracted via OCR,
    and mobile-agent-formatted demos
    are created.
    These demos guide large language models
    (\eg{}, GPT-4)
    to automatically generate structured prompts,
    enabling data agentification.
    The AITW dataset uses manually written instructions directly.

    \item \textbf{Input Formatting:}
    Each sample consists of a screenshot
    and a prompt simulating user intent.

    \item \textbf{Action Annotation:}
    For AITW,
    we extract actions from interaction logs.
    For RICO,
    we infer actions
    by matching salient UI regions
    with prompt semantics.

    \item \textbf{Filtering:}
    We remove samples with low image quality,
    incomplete metadata,
    or ambiguous instructions
    to ensure valid training and evaluation data.
\end{itemize}


\subsection{Evaluation Metrics}
\label{app:metrics}

Beyond the main metrics described in the main text,
we also evaluate the perceptual similarity
between clean and triggered images
using PSNR and SSIM scores,
where higher values indicate that visual perturbations
are less perceptible to users.

\subsection{Fine-tuning Configurations}
\label{app:implementation}

Poison optimization is conducted on frozen VLM backbones
(\eg{}, LLaVA-Mobile)
using the Adam optimizer
with a learning rate of 0.01 and a batch size of 10.
Perturbations are constrained
within an \( \ell_\infty \) bound of \( \epsilon=8.0/255.0 \).
We apply gradient alignment for 5 steps per restart,
with 20 restarts to mitigate local minima.
The poisoning ratio is fixed at 20\%.
Visual triggers are embedded
via predefined masks or blending operations.
Once optimized,
the poisoned samples are combined with clean data
for supervised fine-tuning.
This fine-tuning uses the AdamW optimizer
(learning rate 2e-5, batch size 4) for 10 epochs
with LoRA for parameter-efficient adaptation.
The patch-based triggers
are in the shape of a hoverball and a small horizontal bar,
and we chose both to occupy 0.1\% and 2\% of the screen,
respectively.
The blending rates for Blended trigger is 0.2.
All experiments are conducted on 6 GPUs with 80 GB memory.
Unless otherwise specified,
our main experiments
use the Hoverball trigger,
LLaVA-Mobile as the backbone,
Type III (malicious hijack) attack,
and the RICO dataset.

\subsection{Defense Details}

We standardize the preprocessing steps as follows:

\begin{itemize}
  \item \textbf{Random Crop:} 20\% of the image area is randomly cropped. This step introduces spatial perturbation and tests the spatial robustness of the model.
  \item \textbf{Resize:} The image is resized to 80\% of the original resolution, simulating real-world scaling artifacts.
  \item \textbf{JPEG Compression:} A JPEG quality factor of 50 is applied. This introduces compression noise and tests robustness against low-quality encoding.
\end{itemize}